\documentclass[doublecol]{epl2} 
\usepackage{amsmath}

\title{Athermal Jamming vs. glassy dynamics for particles with exponentially decaying repulsive pair interaction potentials with a cutoff}
\shorttitle{Athermal Jamming vs. glassy dynamics for exponentially decaying interactions}

\author{Nicolas Wohlleben\inst{1} \and Michael Schmiedeberg\inst{1}}
\shortauthor{N. Wohlleben \etal}

\institute{                    
  \inst{1} Institut f\"ur Theoretische Physik I, Friedrich-Alexander-Universit\"at Erlangen-N\"urnberg (FAU), Staudtstra{\ss}e 7, 91058 Erlangen, Germany
}


\abstract{We study athermal jamming as well as the thermal glassy dynamics in systems composed of spheres that interact according to repulsive interactions that exponentially decay as a function of distance. As usual, a cutoff is employed in the simulations. While the athermal jamming transition that is determined by trying to remove overlaps is found to depend on the arbitrary and therefore unphysical choice of the cutoff, we do not find any athermal jamming transition or crossover that only relies on the physical decay length. In contrast, the glassy dynamics mainly depends on the decay length. Our findings constitute another demonstration of the fact that the athermal jamming transition is not related to thermal glassy dynamics. In addition, we argue that interactions without sharp physical cutoff should be considered more often as a model system in jamming. By exploring how widely-used theoretical approaches or methods of analysis in the field of jamming have to be changed in order to not depend on unphysical cutoffs will lead to deeper insights into the nature of athermal and thermal jamming.
  }

\begin{document}

\maketitle

\section{Introduction}

Jamming in athermal systems that are composed of spheres with finite-ranged repulsive interactions can be described and studied by a energy landscape exploration method introduced by O'Hern et al. \cite{OhernLangerLiuNagel,OhernSilbertLiuNagel}. The spheres are initially placed at random positions and then one tries to remove the overlaps by minimizing the energy without crossing energy barriers. In case all overlaps can be removed, the system is called unjammed, while in case remaining overlaps the system is termed jammed. In the large system limit a sharp transition is found from unjammed to jammed packings of spheres \cite{OhernLangerLiuNagel,OhernSilbertLiuNagel} for increasing packing fraction. Close to this athermal jamming transition critical power law scaling of various quantities can be observed (cf. \cite{liureview} as a review).

In thermal systems jamming can be defined in various ways. In many particulate systems one can observe a dramatic slowdown of the dynamics if the density is increased, the temperature is decreased, or external perturbations like shear forces are decreased \cite{liunature}. Here we mainly focus at the slowdown of the dynamics as a function of the packing fraction at small temperatures. In case typical timescales of rearrangements in the systems become larger than the time that is accessible in experiments or simulations such that the system becomes effectively non-ergodic, one often speaks of the (dynamical) glass transition (see, e.g., \cite{berthier11}). Note that the dynamical glass transition usually differs from a possible structural or ideal glass transition that might be defined by the configurational entropy \cite{berthier11,mari09,parisi10,Charbonneau17}.

While it has been conjectured initially that the athermal jamming transition might be the end point of the dynamical glass transition as some kind of thermal jamming transition in the limit of small temperatures \cite{liunature,OhernSilbertLiuNagel}, theoretically they can occur in the same mean field approaches but as distinct phenomena \cite{mari09,parisi10,Charbonneau17}. In simulations the differences between athermal and thermal jamming have been discussed, though they usually occur close together (see, e.g., \cite{ikeda}). 

In this work we study athermal jamming and glassy dynamics for a system with repulsive interactions that decay exponentially with the distance but ideally do not posses a natural cutoff. This model interaction can be seen as simplification of DLVO-interactions that occur in charge-stabilized colloidal systems \cite{dlvo1,dlvo2} or for dust particles in plasmas \cite{plasma}, where the exponential decay is due to screening characterized by the so-called screening length. In simulations the interactions are usually implemented with a cutoff. Here we demonstrate that athermal jamming that is determined with the usual protocol is dominated by the (unphysical) choice of the cutoff, while the glassy dynamics in thermal systems mainly depends on the screening length. While these results might not be considered to be unexpected, we want to stress the importance of studying athermal and thermal jamming in systems with interactions that do not possess a sharp, unique cutoff as many definitions and protocol depend on such a cutoff. We argue that the study of interactions without natural cutoff might help to reveal the true nature of jamming that should not depend on an arbitrary, unphysical cutoff.

\section{Method}
\label{sec:method}

A monodisperse system of $N$ spheres in three spatial dimensions is considered. The interaction between the spheres is purely repulsive and given by the pair potential
\begin{equation}
  V(r)=\begin{cases}\epsilon \exp\left(-r/l_{\textnormal{s}}\right)-V_{\textnormal{c}}(r), &r<l_{\textnormal{c}},\\ 0, &r\geq l_{\textnormal{c}},\end{cases}
\end{equation}
where $r$ is the distance between the particles, $\epsilon$ sets the energy scale, $l_{\textnormal{s}}$ corresponds to the screening lengths, i.e., the physically motivated length scale of the interaction potential, and $l_{\textnormal{c}}$ is a cutoff length that is usually introduced as a purely technical parameter in order to simplify simulations. The function $V_{\textnormal{c}}(r)$ is a linear function that is chosen such that the interaction potential and in case of athermal jamming in addition its derivative do not possess any discontinuity at $r=l_{\textnormal{c}}$.

To determine the jamming behavior at zero temperature, we employ the same energy landscape exploration method as in \cite{OhernLangerLiuNagel,OhernSilbertLiuNagel}: The particles are initially placed at random positions into a simulation box with periodic boundary conditions and volume $V$ and with a given number density $\rho=N/V$. Here $N=700$ particles are used. Note that depending on what length one considers as typical size of the particle, one can define a packing fraction $\phi_{\textnormal{s}}=\pi\rho l_{\textnormal{s}}^3/6$ given by the screening length $l_{\textnormal{s}}$ or $\phi_{\textnormal{c}}=\pi\rho l_{\textnormal{c}}^3/6$ depending on the cutoff length $l_{\textnormal{c}}$. By employing a conjugate gradient method we then minimize the energy, i.e., we determine the local minimum of the energy without crossing any energy barriers. As in \cite{OhernLangerLiuNagel,OhernSilbertLiuNagel} the system is considered to be (athermally) unjammed, if all overlaps can be removed by the minimization. If overlaps prevail, the system is called jammed. To avoid the lengthy final removal of small overlaps we consider configurations to be unjammed if the overlap energy $E$ becomes smaller than $10^{-10}\epsilon$. As a consequence, for counting the number of contacts, overlaps are only counted if larger than $10^{-3}l_{\textnormal{c}}$. We checked that the results do not depend on these choices and that for harmonic interactions the results from \cite{OhernLangerLiuNagel,OhernSilbertLiuNagel} can be reproduced.

In order to explore the dynamics of thermal systems, we perform local Monte Carlo simulations, i.e., Monte Carlo simulations with only local moves such that the number of Monte Carlo sweeps can be used as a approximate measure of the time \cite{Sanz2010}. We determine the relaxation time (number of Monte Carlo sweeps) $\tau$ for which the self-intermediate scattering function for $k=2\pi \phi^{1/3}/l$ decays to $1/e$ of its initial value. For the analysis where the cutoff is taken as length scale, $\phi=\phi_{\textnormal{c}}$ and $l=l_{\textnormal{c}}$. If the screening length is taken in the analysis,  $\phi=\phi_{\textnormal{s}}$ and $l=l_{\textnormal{s}}$. For the dynamical simulations, we employ a bidisperse system, where half of the $N=1000$ particles are larger by a factor $1.4$, i.e., the same interaction potential as for athermal jamming is employed, but all lengths are scaled by a factor $1.4$ in case of two large particles and by $1.2$ if a small particle interacts with a large one. We start with a packing fraction $\phi_{\textnormal{c}}$ or $\phi_{\textnormal{s}}=0.05$ and later increase the packing fraction in intervals of $0.05$. At each density, the relaxation time is measured after an initial relaxation of $10^5$ Monte Carlo sweeps with adaptive maximum displacements per step. Afterwards the relaxation time is measured with a constant maximum size of displacements per Monte Carlo step of $l/\phi^{1/3}/20$. Different temperatures from $k_{\textnormal{B}}T/\epsilon=0.1$ to $k_{\textnormal{B}}T/\epsilon=0.5$ are considered.

\section{Results}

\subsection{Athermal Jamming}

\begin{figure}[htb]
\onefigure[width=\columnwidth]{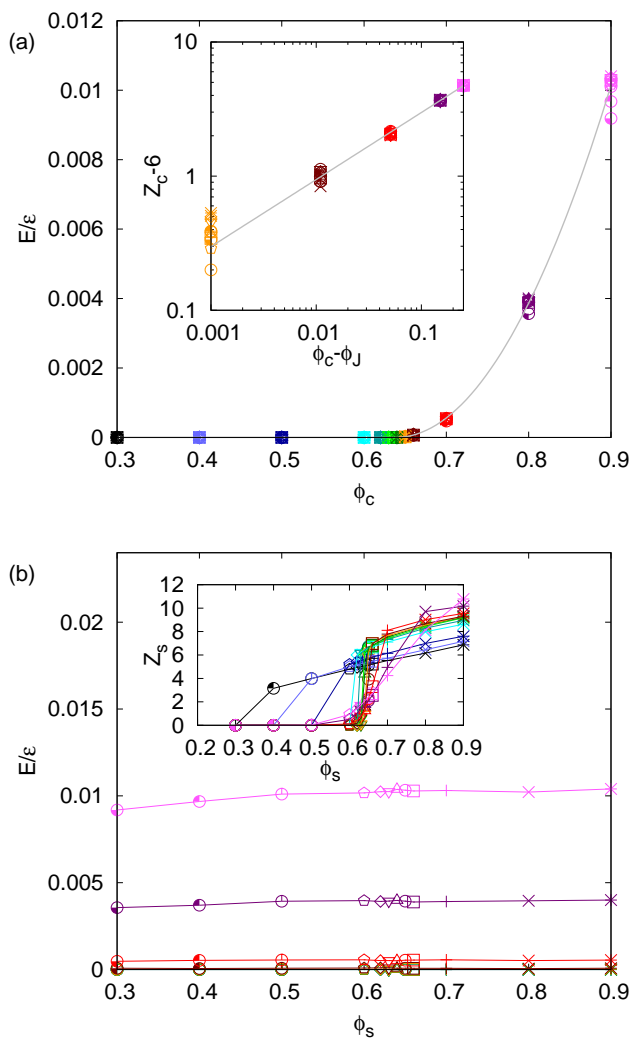}
\caption{Overlap energy per particle after the minimization for various cutoff lengths and screening lengths that are denoted by the corresponding packing fractions $\phi_{\textnormal{c}}$ and $\phi_{\textnormal{s}}$, respectively. In (a) the overlap energy is shown as a function of the packing fraction $\phi_{\textnormal{c}}$ based on the cutoff length $l_{\textnormal{c}}$, while in (b) the same data is plotted as a function of $\phi_{\textnormal{s}}$ based on the screening length $l_{\textnormal{s}}$. In all figures the color denotes $\phi_{\textnormal{c}}$ as can be read of in (a) and in addition is shown in the legend of fig.~\ref{fig.2}(b) while the symbols mark $\phi_{\textnormal{s}}$ as visible in (b) and in the legend of fig.~\ref{fig.2}(a). The grey line in the main figure of (a) is a quadratic fit function that starts from zero at a transition packing fraction $\phi_{\textnormal{J}}$. From the fit, we find $\phi_{\textnormal{J}}=0.638$. The inset in (a) shows the well-known power-law scaling with exponent 0.5 for the number of contacts $Z_{\textnormal{c}}$ per particle, defined of overlaps concerning the cutoff length $l_{\textnormal{c}}$, in excess of the isostatic contact number 6 as a function of the packing fraction $\phi_{\textnormal{c}}$ above the transition packing fraction $\phi_{\textnormal{J}}$. In the inset in (b) $Z_{\textnormal{s}}$, which gives the number of neighbors per particle that are closer than the screening length $l_{\textnormal{s}}$, is plotted as a function of $\phi_{\textnormal{s}}$.}
\label{fig.1}
\end{figure}

In fig.~\ref{fig.1} the overlap energy that is reached after a minimization towards a local minimum, i.e., without crossing energy barriers is shown. In fig.~\ref{fig.1}(a) it is plotted as a function of $\phi_{\textnormal{c}}$. In fig.~\ref{fig.1}(b) the same data is shown as a function of $\phi_{\textnormal{s}}$. Therefore, in fig.~\ref{fig.1}(a) the cutoff length $l_{\textnormal{c}}$ determines the packing fraction while in fig.~\ref{fig.1}(b) the data is organized by the screening length $l_{\textnormal{s}}$.

In fig.~\ref{fig.1}(a) the well known behavior with no overlaps below a transition packing fraction $\phi_{\textnormal{J}}$ and an increasing overlap energy above the transition is observed. The transition occurs at the same $\phi_{\textnormal{J}}$ for all screening lengths $l_{\textnormal{s}}$ and is in agreement with the transition packing fraction of spheres with harmonic interactions as reported in \cite{OhernLangerLiuNagel,OhernSilbertLiuNagel}. The inset of fig.~\ref{fig.1}(a) shows the difference of the number of per particle contacts $Z_{\textnormal{c}}$, i.e., neighbors that are closer than the cutoff length $l_{\textnormal{c}}$, and the isostatic contact number 6 as a function of the difference of the packing fraction $\phi_{\textnormal{c}}$ and the transition packing fraction $\phi_{\textnormal{J}}$. For large packing fractions $\phi_{\textnormal{c}}$ the data depends slightly on $l_{\textnormal{s}}$, but otherwise $l_{\textnormal{s}}$ hardly affects the overlap energy.

If the overlap energy is plotted as a function of the packing fraction $\phi_{\textnormal{s}}$ that is given by the screening length $l_{\textnormal{s}}$, all curves with $\phi_{\textnormal{c}}\leq \phi_{\textnormal{J}}$ are zero everywhere, and the curves for $\phi_{\textnormal{c}}>\phi_{\textnormal{J}}$ are slightly increasing with increasing $\phi_{\textnormal{s}}$. Otherwise, there is no indication for any transition or crossover that could explain a connection between the rapid slowdown of the dynamics that we present in the next subsection and the inherent structures whose overlap energy is shown here.

The inset of fig.~\ref{fig.1}(b) shows the number of contacts $Z_{\textnormal{s}}$ given by the number of neighbors that are closer than the screening length $l_{\textnormal{s}}$ as a function of $\phi_{\textnormal{s}}$. A sharp increase only occurs in case $\phi_{\textnormal{s}}\approx \phi_{\textnormal{c}}\approx \phi_{\textnormal{J}}$. For $\phi_{\textnormal{c}}<\phi_{\textnormal{J}}$ the number of contacts $Z_{\textnormal{s}}$ start to deviate from 0 below $\phi_{\textnormal{J}}$ while for $\phi_{\textnormal{c}}>\phi_{\textnormal{J}}$ there is a continuous increase that roughly starts at $\phi_{\textnormal{J}}$. As a consequence, no universal transition behavior can be found if $l_{\textnormal{s}}$ is used for the analysis of contacts.

\begin{figure}[htb]
\onefigure[width=\columnwidth]{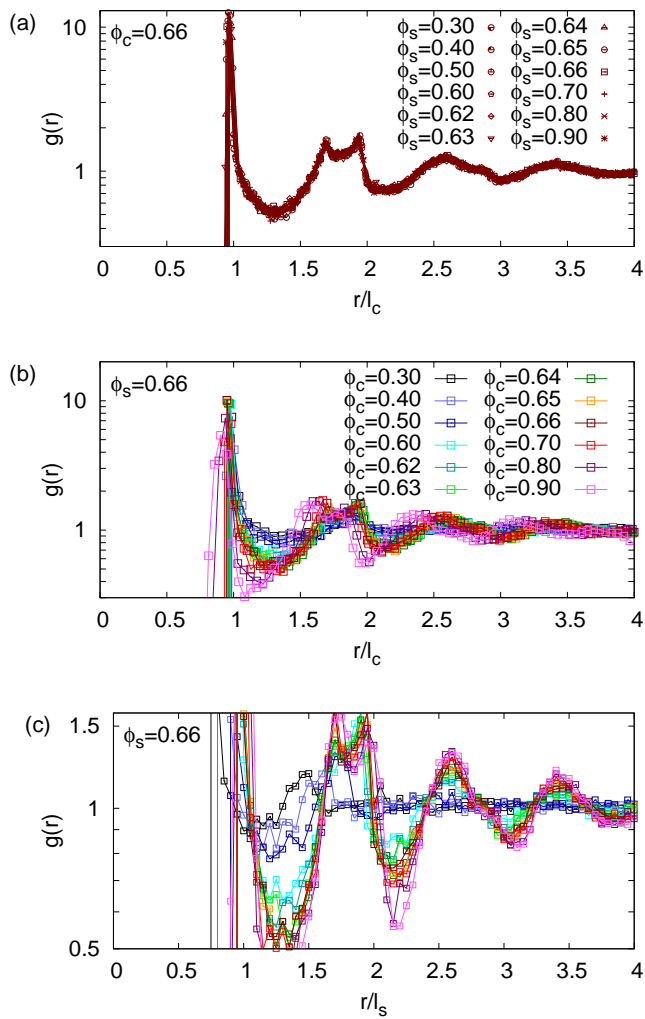}
\caption{Pair distribution function of the structures obtained after minimizing the overlap energy without crossing energy barriers. The length is plotted in units of (a,b) the cutoff length $l_{\textnormal{c}}$ or (c) the screening length $l_{\textnormal{c}}$. (a) All curves are for the same $\phi_{\textnormal{c}}=0.66$ and various $\phi_{\textnormal{s}}$. (b,c) $\phi_{\textnormal{s}}$ is kept constant at 0.66 and $\phi_{\textnormal{c}}$ is varied. The colors and symbols, as also shown in the legend, are the same as in fig.~\ref{fig.1}.}
\label{fig.2}
\end{figure}

In fig~\ref{fig.2} we present pair distribution functions $g(r)$ for the structures obtained after the minimization process. In fig.~\ref{fig.2}(a) the length is plotted in units of the cutoff length $l_{\textnormal{c}}$ and curves slightly above the athermal jamming transition, i.e., for constant $\phi_{\textnormal{c}}=0.66$, are shown for various $\phi_{\textnormal{s}}$. All curves collapse onto each other within the size of the symbols. Therefore, close to athermal jamming, $g(r)$ is almost independent of the screening length $l_{\textnormal{s}}$. As a consequence, the inherent structures close to athermal jamming depend on the cutoff length $l_{\textnormal{c}}$ but not on the screening length $l_{\textnormal{s}}$.

In fig.~\ref{fig.2}(b,c) $g(r)$ for a constant $\phi_{\textnormal{s}}=0.66$ and various $\phi_{\textnormal{c}}$ are shown. Therefore, cases further away from the athermal jamming transition are included. If the length is plotted in units of the cutoff length $l_{\textnormal{c}}$ as depicted in fig.~\ref{fig.2}(b), one finds that the first peak of $g(r)$ occurs at $l_{\textnormal{c}}$ but otherwise the other features of $g(r)$ are not determined by $l_{\textnormal{c}}$. If the length is shown in units of $l_{\textnormal{s}}$ (see fig.~\ref{fig.2}(c)), there are still significant differences between the curves. However, for sufficiently large density ($\phi_{\textnormal{c}}\geq 0.6$) all peaks with the exception of the first one occur at the same lengths. As a consequence, while the first peak is given by the cutoff, otherwise the structure is dominated by the screening length. As we will show in the next subsection, the dynamics is mainly given by the screening length and therefore we conclude that the dynamics hardly depends on the contact behavior, but mainly is given by longer-ranged correlations.

\subsection{Glassy dynamics of thermal systems}

The thermal jamming of particles that interact according to the DLVO-potential has been studied, e.g., in \cite{jamDLVO}. Here we want to explore how the two length scales, i.e., the screening length $l_{\textnormal{s}}$ and the cutoff $l_{\textnormal{c}}$, affect the glassy dynamics.

\begin{figure}[htb]
\onefigure[width=\columnwidth]{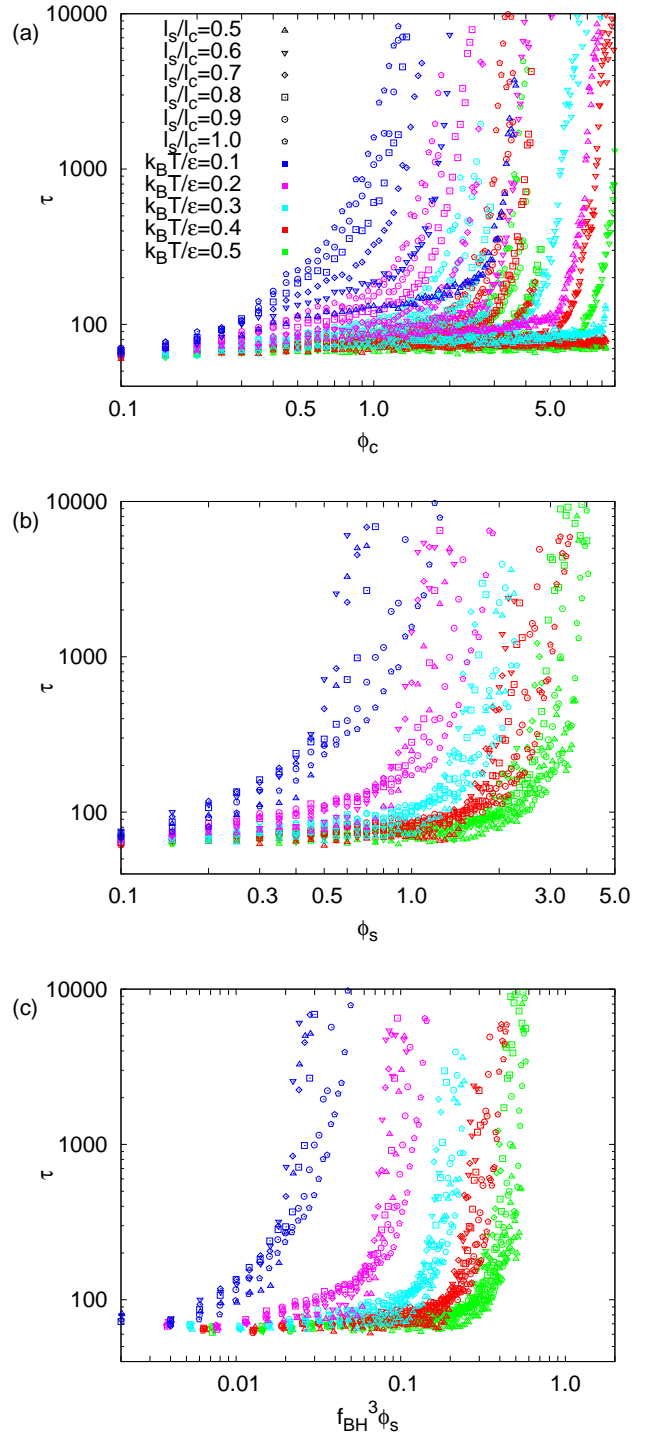}
\caption{Relaxation time from dynamical Monte Carlo simulations as function of (a) $\phi_{\textnormal{c}}$, (b) $\phi_{\textnormal{s}}$, and (c) $\phi_{\textnormal{s}}$ rescaled with a factor based on the Barker-Hendersen approximation \cite{barker} as explained in the text. The temperature is denoted by the color while the symbols indicate the ratios of the screening length $l_{\textnormal{s}}$ and the cutoff length $l_{\textnormal{c}}$.}
\label{fig.3}
\end{figure}

In fig.~\ref{fig.3} the relaxation time obtained by dynamical Monte-Carlo simulations as described in sec.~\ref{sec:method} is plotted as function of the packing fraction. If plotted as function of $\phi_{\textnormal{c}}$ and for an analysis based on the cutoff length $l_{\textnormal{c}}$ (see fig.~\ref{fig.3}(a)), the slowdown of the dynamics for different ratios of $l_{\textnormal{s}}/l_{\textnormal{c}}$ occurs at very different packing fractions. Obviously, the cutoff length is not a good way to characterize the dynamics.

As a function of $\phi_{\textnormal{s}}$ and with an evaluation also otherwise based on the screening length $l_{\textnormal{s}}$ (see fig.~\ref{fig.3}(b)), the curves arrange in groups where the slowdown for a given temperature (as indicated by the color) is at least similar even for different $l_{\textnormal{s}}/l_{\textnormal{c}}$. As a consequence, the dynamics of the system mainly depends on the screening length $l_{\textnormal{s}}$. The cutoff length still has an significant influence on the dynamics but in contrast to the athermal jamming cannot be used as control parameter alone.

These results probably are not surprising: Concerning the dynamics in thermal systems, one usually does not expect that the thermal dynamics strongly depends on a cutoff length. For example, Yukawa-like particles in plasmas can be studied by using the isomorph invariance \cite{dyre1} that relies on the screening lengths. Furthermore, there are various way to approximately map the dynamics of soft particle onto the dynamics of hard spheres \cite{schmiedeberg2011,medina2011,maiti2018b} and for the employed methods most cutoffs are negligible: An effective diameter that is determined according to the Barker-Handersen \cite{barker} or the Andersen-Weeks-Chandler method \cite{andersen1971} is hardly affected by a change of the cutoff if the cutoff length is sufficiently large. As a consequence, from the mapping-approaches one indeed expect that glassy dynamics does not strongly depend on the cutoff.

Note that in fig.~\ref{fig.3}, short cutoff lengths were chosen to demonstrate their (weak) influence. We can correct the packing fraction $\phi_{\textnormal{s}}$ by employing a factor $f_{\textnormal{BH}}$ that is given as ratio of the Barker-Hendersen length \cite{barker} of the interaction potential without cutoff and the one of the potential with cutoff. In fig.~\ref{fig.3}(c) we show that this rescaling of the packing fraction reduces the dependence on the cutoff. Probably a further reduction of the cutoff-dependence with improved rescalings as, e.g., in \cite{andersen1971,schmiedeberg2011,medina2011}.

Finally, we want to note that multiple reentrant glass transitions are observed in ultrasoft systems for increasing overlaps \cite{berthier10,schmiedeberg13,miyazaki16}. Obviously this dynamics is not connected at all to the conventional definition of athermal jamming where only the difference between systems with overlaps and systems without overlaps matters.

\section{Discussion and Conclusions}

We studied the athermal jamming behavior for particles with exponentially decaying repulsive interaction potentials and a cutoff. We find that according to the conventional definition of jamming the athermal jamming transition depends on the cutoff length but not on the screening length. No other transition or crossover is observed in athermal systems if the screening length is taken to define the packing fraction. In contrast, the slowdown of the thermal glassy dynamics as a function of packing fraction mainly depends on the screening lengths.

Concerning the connection of structure and glassy dynamics our results suggest that the dynamics mainly depends on the longer-ranged correlations while the structure close to contact is less important. Note that in our system the repulsive forces close to contact are small and therefore we are not close to the hard-sphere limit.

On a first view the results might appear trivial. The conventional definition of the athermal jamming transition relies on the range of the interaction potential. However, we want to point to important, non-trivial conclusions and questions that arise due to our findings:

Sometimes the athermal jamming transition has been described as some kind of end point of a thermal jamming line that is related to the glass transition \cite{OhernSilbertLiuNagel,liunature}. For the system that we consider here such an interpretation is impossible, because the athermal jamming transition depends on the cutoff while the glass transition is mainly controlled by the screening length. Athermal and thermal jamming can occur at very different packing fractions. Therefore, by using the exponential interaction potential with a cutoff athermal and thermal jamming can be tuned and studied separately.

Since the athermal jamming transition in our system can be seen as an artifact caused by basing the definition on a cutoff, there is an imminent question: Is there an athermal jamming transition (or at least a crossover) at all for interaction potentials that are not finite-ranged? Obviously, mechanical properties like elastic modules of a suspension depend on the density. Our results suggest that the rigidification for increasing density is a smooth process as there is no obvious physical way to define whether there is a contact between neighboring particles or not. As a consequence, a term like isostaticity should not be used for particle where the interaction is not finite-ranged. 

In many theoretical works (see, e.g., \cite{mari09,parisi10,Charbonneau17}) the differences between athermal jamming and the glass transition are carefully considered. However, the jamming line that occurs as limiting line of the glassy behavior predicted by mean-field theory (cf., \cite{parisi10,Charbonneau17}) probably does not exist at all without cutoff. As we have shown by using the exponential interactions with cutoff, the dynamics glass transition and athermal jamming can be tuned almost independently. Therefore we suggest that it could be of great interest to use this model system to explore how other predicted intermediate transitions behave, e.g., the Gardner transition.

Many simulation studies dealing with the thermal glass transition assume or find that the thermal glass transition occurs close to athermal jamming (see, e.g., \cite{berthier11,BerthierandWitten,BerthierandWitten1,haxton2011,ikeda,artiaco20}). In addition, many works successfully employ an analysis that is based on an athermal jamming-like definition with overlaps \cite{nagel05,corwin17,maiti2018,maiti2018c} or deal with inherent structures (in the sense of local minima of the energy landscape) \cite{stilinger95,sastry02,ozawa12,scalliett19,speck2020,artiaco20} though without cutoff the energy landscape might consist of only one connected energy basin just as in systems with cutoff below athermal jamming \cite{maiti2018}. However, in systems where athermal jamming is an artifact related to an arbitrary cutoff, why should the athermal jamming point be a valid starting point to explore glassy dynamics or to find properties of the glass transition? Is there something special about the often employed harmonic or Hertzian interactions that athermal and thermal jamming seem to be closely related in these systems?

We believe that it is of great importance to find out which properties connected to glassy behaviors rely on a cutoff lengths or not. The definitions, analysis, or theoretical approaches should be changed such that they are independent of unphysical cutoffs. The interaction potential used in this letter is suitable to explore which properties and phenomena depend on the screening length and therefore can be related to the glassy dynamics and which one rely on the cutoff.

For example, in \cite{milz2013,maiti2018,maiti2018c} we have found an directed percolation transition in time that breaks ergodicity and therefore corresponds to the dynamical glass transition between ergodic states that can reach the ground state and non-ergodic states where the ground state is inaccessible \cite{maiti2018,maiti2018c}. In principle, such an ergodicity breaking can also occur for particles without finite-ranged interactions though the recognition of ground states is more difficult because they no longer can be associated to configurations without overlaps. Therefore, in a future work, we want to study ergodicity breaking without relying on athermal jamming or any definition based on overlaps.

Furthermore, there are many protocols that investigate the energy or free energy landscape, e.g., to study its hierarchical properties close to the glass transition or the Gaardner transition \cite{scalliett19,dennis20}. It is an open question whether the same results can be obtained in absence of a nearby athermal jamming transition and therefore whether these findings are really related to the glass transition.

Finally, in future works we want to find out how jamming in case of additional attractive interactions can be properly defined. In case of short finite-ranged interactions with weak attractions, athermal jamming can be studied in a similar way as for finite-ranged repulsive interactions \cite{koetze18}. However, in colloidal gel-forming systems the interactions and dynamics are more complex and usually at longer distances there is a repulsive interaction as considered in this letter \cite{gel,kohl,kohl17,rp1} and it is not expected that a cutoff length significantly affects the properties of a gel. However, it remains an open question how the mechanical solidification of a gel might be related to the athermal jamming, especially because gelation sometimes is associated to a rigidity percolation transition \cite{rp1,rp2} that is based on whether strands of isostatic particles, as defined in athermal jamming with some cutoff, percolate or not.

\acknowledgments
The project was supported by the Deutsche Forschungsgemeinschaft (Grant No. Schm 2657/3-1).


\begin{thebibliography}{0}







\bibitem{OhernLangerLiuNagel}
\Name{O'Hern C.S., Langer S.A., Liu A.J. \and Nagel S.R.}
   \REVIEW{Phys. Rev. Lett.} {88} {2002} {075507}

\bibitem{OhernSilbertLiuNagel}
\Name{O'Hern C.S., Silbert L.E., Liu A.J. \and Nagel S.R.}
   \REVIEW{Phys. Rev. E}  {68} {2003} {011306}

\bibitem{liureview}
  \Name{Liu A.J. \and Nagel S.R.}
  \REVIEW{Annual Review of Condensed Matter Physics}{1}{2010}{347}
  
\bibitem{liunature}
  \Name{Liu A.J. \and Nagel S.R.}
  \REVIEW{Nature}{396}{1998}{21}
  
\bibitem{berthier11}
  \Name{Berthier L. \and Biroli G.}
  \REVIEW{Rev. Mod. Phys.}{83}{2011}{587}

\bibitem{mari09}
  \Name{Mari R., Krzakala F. \and Kurchan J.}
\REVIEW{Phys. Rev. Lett.}{103}{2009}{025701}

\bibitem{parisi10}
  \Name{Parisi G. \and  Zamponi F.}
  \REVIEW{Rev. Mod. Phys.}{82}{2010}{789}

\bibitem{Charbonneau17}
\Name{Charbonneau P., Kurchan J., Parisi G., Urbani P. \and Zamponi F.}
\REVIEW{Annu. Rev. Condens. Matter Phys.}{8}{2017}{265} 

\bibitem{ikeda}
    \Name{Ikeda A., Berthier L. \and Sollich P.}
   \REVIEW{Phys. Rev. Lett.}{109}{2012}{018301}

\bibitem{dlvo1}
  \Name{Derjaguin B.V. \and Landau L.}
  \REVIEW{Acta Physicochimica (USSR)}{14}{1941}{633}
  
\bibitem{dlvo2}
  \Name{Verwey E.J. \and Overbeek J.T.G.}
  \REVIEW{Theory of the Stability of Lyophobic Colloids (Elsevier, Amsterdam)}{}{1948}

\bibitem{plasma}
  \Name{Ivlev A.V., L\"owen H., Morfill G.E. \and Royall C.P.}
  \REVIEW{Complex plasmas and colloidal dispersions: particle-resolved studies of classical liquids and solids (World Scientific)}{}{2012}

  
\bibitem{Sanz2010}
\Name{Sanz E. \and Marenduzzo D.}
\REVIEW{J. Chem. Phys}{132}{2010}{194102}

\bibitem{jamDLVO}
  \Name{Kumar A. \and Wu J.}
   \REVIEW{Appl. Phys. Lett.}{84}{2004}{4565}

\bibitem{dyre1}
  \Name{Lucco Castello F., Tolias P., Hansen S.J. \and Dyre J.C.}
  \REVIEW{Physics of Plasmas} {26} {2019} {053705}


   
\bibitem{schmiedeberg2011}
\Name{Schmiedeberg M., Haxton T.K., Nagel S.R. \and Liu A.J.}
\REVIEW{EPL} {96} {2011} {36010}

\bibitem{medina2011}
  \Name{Ram\'irez-Gonz\'alez P.E., L\'opez-Flores L., Acu\~na-Campa H. \and Medina-Noyola M.}
  \REVIEW{Phys. Rev. Lett.} {107} {2011} {155701}

\bibitem{maiti2018b}
\Name{Maiti M. \and Schmiedeberg M.}
   \REVIEW{Journal of Physics: Condensed Matter}{31}{2019}{165101}

\bibitem{barker}
  \Name{Barker J.A. \and Henderson D.}
  \REVIEW{Phys. Rev.  A}{1}{1970}{1266}
  
\bibitem{andersen1971}
  \Name{Andersen H.C., Weeks J.D. \and Chandler D.}
  \REVIEW{Phys. Rev. A} {4} {1971} {1597}

\bibitem{berthier10}
\Name{Berthier L., Moreno A. J. \and Szamel G.}
   \REVIEW{Phys. Rev. E} {82} {2010} {060501(R)}
  
\bibitem{schmiedeberg13}
\Name{Schmiedeberg M.}
   \REVIEW{Phys. Rev. E} {87} {2013} {052310}

 \bibitem{miyazaki16}
\Name{Miyazaki R., Kawasaki T. \and Miyazaki K.}
   \REVIEW{Phys. Rev. Lett.} {117} {2016} {165701}

\bibitem{BerthierandWitten}
  \Name{Berthier L. \and Witten T. A.}
  \REVIEW{Europhys. Lett.} {86} {2009} {10001}

\bibitem{BerthierandWitten1}
  \Name{Berthier L. \and Witten T.A.}
  \REVIEW{Phys. Rev. E} {80} {2009} {021502}

\bibitem{haxton2011}
  \Name{Haxton T.K., Schmiedeberg M., \and Liu A.J.}
  \REVIEW{Phys. Rev. E} {83}{2011}{031503}

\bibitem{artiaco20}
   \Name{Artiaco, C., Baldan, P., \and Parisi G.}
   \REVIEW{Phys. Rev. E}{101}{2020}{052605}
   
  \bibitem{nagel05}
   \Name{Corwin, E.I., Jaeger, H.M. \and Nagel, S.R.}
\REVIEW{Nature}{435}{2005}{1075}

\bibitem{corwin17}
   \Name{Morse, P.K. \and Corwin, E.I.}
\REVIEW{Phys. Rev. Lett.}{119}{2017}{118003}

\bibitem{maiti2018}
\Name{Maiti M. \and Schmiedeberg M.}
   \REVIEW{Scientific Reports} {8} {2018} {1837}

\bibitem{maiti2018c}
\Name{Maiti M. \and Schmiedeberg M.}
   \REVIEW{Eur. Phys. J. E}{42}{2019}{38}

\bibitem{stilinger95}
  \Name{Stilinger, F.H.}
  \REVIEW{Science}{267}{1995}{1935}

\bibitem{sastry02}
  \Name{Sastry, S.}
  \REVIEW{Phase Transitions}{75}{2002}{507}
  
\bibitem{ozawa12}
  \Name{Ozawa M., Kuroiwa T., Ikeda A., and Miyazak K.}
  \REVIEW{Phys. Rev. Lett.}{109}{2012}{205701}
  
\bibitem{scalliett19}
\Name{Scalliet C., Berthier L. \and Zamponi  F.}
\REVIEW{Nature Communications}{10}{2019}{5102}

\bibitem{speck2020}
  \Name{Royall C.P., Turci F. \and Speck T.}
  \REVIEW{Dynamical Phase Transitions and their Relation to Structural andThermodynamic Aspects of Glass Physics}{}{2020}{ArXiv:2003.03700}

\bibitem{milz2013}
\Name{Milz L. \and Schmiedeberg M.}
\REVIEW{Phys. Rev. E} {88} {2013} {062308}

\bibitem{dennis20}
  \Name{Dennis R.C. \and Corwin E.I.}
  \REVIEW{Phys. Rev. Lett.}{124}{2020}{078002}


\bibitem{koetze18}
\Name{Koeze D.J. \and Tighe B.P.}
\REVIEW{Phys. Rev. Lett.}{121}{2018}{188002}

  
  
\bibitem{gel}
\Name{van Doorn J.M., Bronkhorst J., Higler R., van de Laar T. \and Sprakel J.}
   \REVIEW{Phys. Rev. Lett.} {118} {2017} {188001}

\bibitem{kohl}
\Name{Kohl M., Capellmann R.F., Laurati M., Egelhaaf S.U. \and Schmiedeberg M.}
\REVIEW{Nature Communications} {7} {2016} {11817}

  
\bibitem{kohl17}
\Name{Kohl M. \and Schmiedeberg M.}
\REVIEW{Eur. Phys. J. E} {40}{2017}{71}


\bibitem{rp1}
 \Name{Tsurusawa H., Leocmach M., Russo J. \and Tanaka H.}
\REVIEW{Science Advances}{5}{2019}{eaav6090}

\bibitem{rp2}
  \Name{Zhang S., Zhang L., Bouzid M., Zeb Rocklin D., Del Gado E. \and Mao X.}
\REVIEW{Phys. Rev. Lett.}{123}{2019}{058001}

\end{thebibliography}
\end{document}